\documentclass[twocolumn,preprintnumbers,amsmath,amssymb,superscriptaddress,prl]{revtex4}

\usepackage[T1]{fontenc}
\usepackage{lmodern}
\usepackage[utf8]{inputenc}
\usepackage[english]{babel}
\usepackage{graphicx}
\usepackage{dcolumn}
\usepackage{bm}
\usepackage{color}
\usepackage{soul}
\usepackage[normalem]{ulem}
\usepackage{booktabs}
\usepackage{hhline}

\makeatletter
\def\uwave{\bgroup \markoverwith{\lower3.5\p@\hbox{\sixly \textcolor{red}{\char58}}}\ULon}
\font\sixly=lasy6 
\makeatother

\begin{document}

\title{Pressure induced crossing of the core-levels in 5d metals}

\author{Alexey A. Tal}
\email{aleta@ifm.liu.se}
\affiliation{Theory and Modeling, IFM-Material Physics, Linköping University, SE-581 83, Linköping, Sweden}
\affiliation{Materials Modeling and Development Laboratory, National University of Science and Technology 'MISIS', 119049, Moscow, Russia}  


\author{Mikhail I. Katsnelson}
\affiliation{Radboud University of Nijmegen, Institute for Molecules and Materials, Heyendaalseweg 135, 6525AJ Nijmegen, The Netherlands}%
\affiliation{Department of Theoretical Physics and Applied Mathematics, Ural Federal University, Mira str, 19, Ekaterinburg, 620002, Russia}%

\author{Marcus Ekholm}
\affiliation{Theory and Modeling, IFM-Material Physics, Linköping University, SE-581 83, Linköping, Sweden}

\author{Johan Jönsson}
\affiliation{Theory and Modeling, IFM-Material Physics, Linköping University, SE-581 83, Linköping, Sweden}

\author{Leonid Dubrovinsky}
\affiliation{Bayerisches Geoinstitut, Universitat Bayreuth, D-95440 Bayreuth, Germany}
\author{Natalia Dubrovinskaia}
\affiliation{Bayerisches Geoinstitut, Universitat Bayreuth, D-95440 Bayreuth, Germany}

\author{Igor A. Abrikosov}
\affiliation{Theory and Modeling, IFM-Material Physics, Linköping University, SE-581 83, Linköping, Sweden}%
\affiliation{Materials Modeling and Development Laboratory, National University of Science and Technology 'MISIS', 119049, Moscow, Russia}  

\begin{abstract}
Pressure induced interaction between core electrons, the core level crossing (CLC) transition has been observed in hcp Os at P~400 GPa [L. Dubrovinsky, et al., Nature 525, 226–229 (2015)]. In this work, we carry out a systematic study of the influence of pressure on the electronic structure in all metals of the 5d series (Hf,Ta,W,Re,Os,Ir,Pt,Au) using first-principles electronic structure calculations. We have found that CLC is a general effect for this series of metals. While in Pt it occurs at ~1500 GPa, at a pressure substantially higher than in Os, in Ir it occurs already at 80 GPa. Moreover, we predict that in Re the CLC transition may appear at ambient pressure. We analyze the shifts of the CLC transition pressure across the series within the Thomas-Fermi model, and show that the effect has many common features to the atomic collapse in the rare-earth elements.
\end{abstract}

\maketitle

The properties of matter are determined by its electronic structure, which is sensitive to external parameters like pressure, temperature, and chemical composition. A modification of the electronic structure may lead to a phase transition and a change the material properties, allowing for a synthesis of new materials. The behavior of matter at extreme conditions has always drawn attention of a broad research community. For instance, compression may result in qualitative changes of the state of a solid such as structure, magnetic state and conductivity.
The metals of 5d group of periodic table are of particular interest for high-pressure study due to their remarkable properties. Hafnium (Hf) has attracted great scientific and technological interest due to the position of its d-band in the middle of a broad sp band, which has an impact on its electronic and superconducting properties \cite{Duthie1977,Skriver1985,Gyanchandani1999,Vohra2001}. Instability of hcp phase of Hf at high pressure was proven both theoretically and experimentally \cite{Hrubiak2012,Hao2011}.  Tantalum is stable even at the pressure of hundreds of GPa \cite{Moriarty2002,Cynn1999}. Even though the stability at higher pressure is still under debate \cite{Yao2013,Burakovsky2010}, only hcp phase was observed experimentally. Ta shows high chemical and thermodynamical stability, having a melting point at ambient pressure of about 3950K \cite{Dewaele2010} and it is widely used in the microelectronics industry for producing integrated circuits. The strength of tungsten are of considerable importance for optimizing the design and operation of high-pressure apparatus. Pressure calibration in diamond anvil cells is largely based on equations of state derived from shock data for standard materials such W, Mo, Cu \cite{Dewaele2004,Chijioke2005}. Neither theoretical calculations nor experimental observations suggest that W may suffer a structural transition under pressure. Rhenium (Re) was studied under extremely high pressures up to 600~GPa and showed no structural transformations in the considered pressure range. That makes Re a good candidate for  ultra-high pressure calibration\cite{Anzellini2014,Zha2004,Dubrovinsky2012,Verma2003}. 
In the recent paper Dubrovinsky et al. \cite{Igor2015} have presented experiments on osmium compression up to over 770~GPa in diamond anvil-cell. Even though hcp phase turned out to be stable in the whole pressure range, authors have discovered two peculiarities in c/a  behavior under pressure. The first one appeared at \textasciitilde 150~GPa and was attributed to the so-called electronic topological transition (ETT) \cite{Lifshitz}, while the second one at \textasciitilde 440~GPa was explained by the crossing of core-levels (CLC). The CLC is a novel type of the electronic transition,  and requires systematic investigation in order to obtain full understanding of its nature. 
A stability of the fcc iridium under pressure has been debated for years. A formation of the complex superlattice in iridium under pressure of  59~GPa has been reported \cite{Cerenius2000}, however, in other experimental and theoretical studies such a structure has not been observed \cite{Grussendorff2003}. Platinum is also widely used as a pressure standard and is known to be stable in the fcc structure up to 600~GPa \cite{Ono2011,Belonoshko2012}. The uniqueness of gold and its important role in modern science is closely related to its exceptional stability to chemical reactions, extreme pressures and temperatures \cite{Hammer1995,Batani2000}. Gold in the fcc structures become unstable in favor of the hcp structure only under pressure of 240~GPa and at elevated temperatures.\cite{Dubrovinsky2007}. Thus these metals show structural stability under high pressure, they are non-magnetic and may be good candidates for investigation of the transitions of another kind.

It is well-known that chemical bonding in solids is mainly due to valence electrons while core-electrons are often considered "frozen" and do not influence macroscopic properties of the materials. However, strong rearrangements of 5p and 4f states at the CLC transition may affect the valence electrons due to non-local nature of the electron interactions and therefore could indirectly influence the structural properties as shown in \cite{Igor2015}. 
In this Letter we analyze the behavior of the core-levels in 5d-metals under pressure by means of ab initio calculations and study the influence of the structure on the core-levels interplay.

For our calculations we used WIEN2k code \cite{wien2k} with LDA functional and k-mesh of 32x32x32 k-points in fcc and bcc structures and 39x39x21 in hcp structure. The radius of real-space muffin-tin sphere varied under different pressures, while the product $K_{max} \cdot R_{MT}$ was kept equal to 10. The spin-orbit interaction is included variationally.

\begin{figure*}[]
\begin{flushleft}

  \begin{tabular}{@{}cc@{}}
    \includegraphics[width=0.45\textwidth]{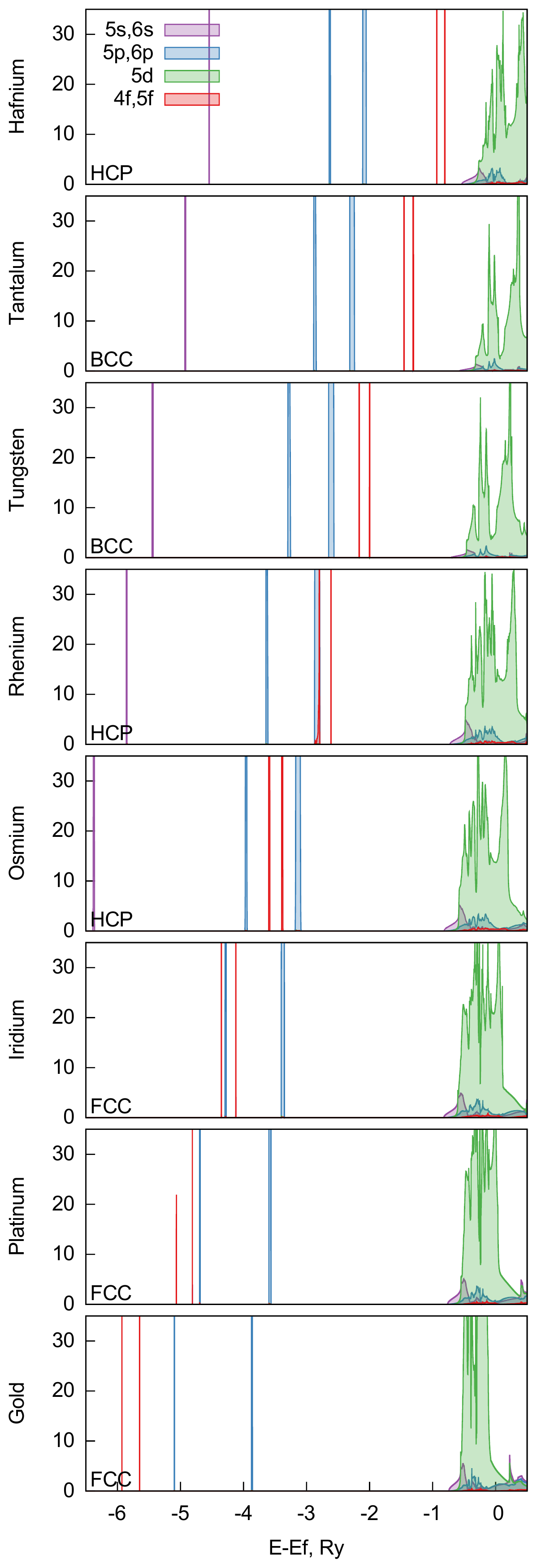} &
    \includegraphics[width=0.45\textwidth]{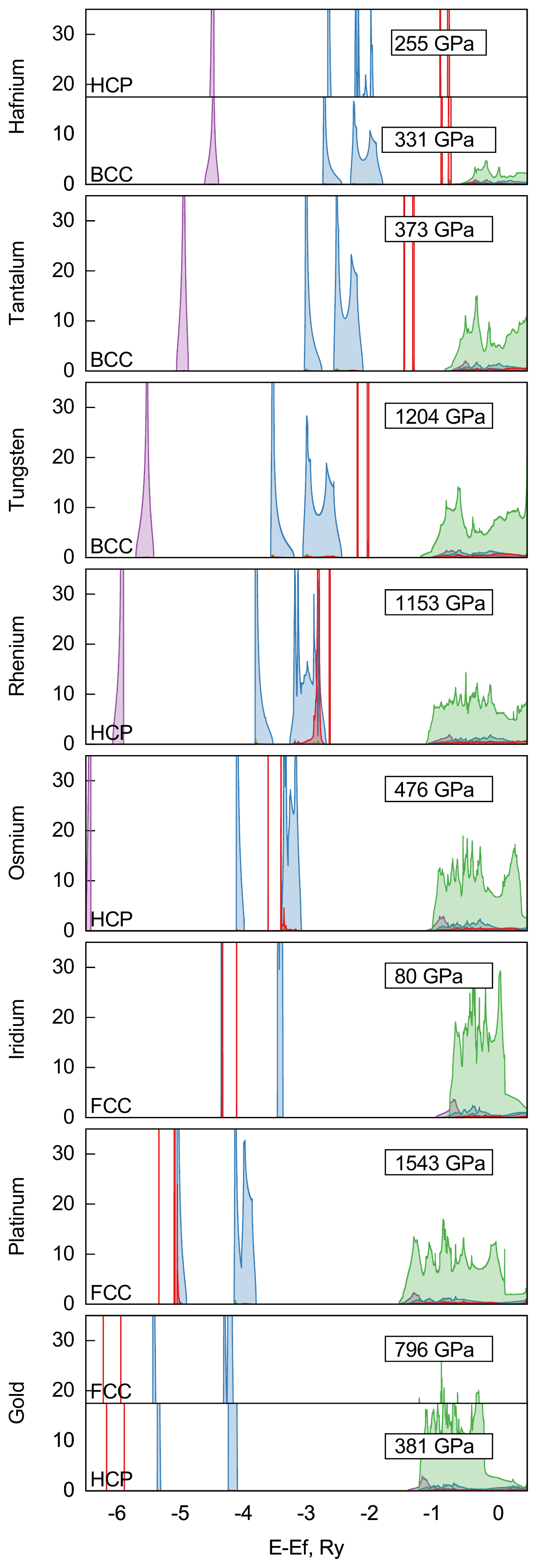}  \\

  \end{tabular}
  \caption{(left) Electronic density of states (DOS) for 5d-metals from Hf to Au at zero pressure; (right) DOS for 5d metals from Hf to Au  in structures experimentally known to be stable at high pressure. For each element the pressure shown in the figure correspond either to the pressure of the core level crossing transition or to the highest pressure considered in this study.}
\label{fig:fcc}
\end{flushleft}
\end{figure*}

Calculations of the density of states (DOS) have been carried out in 5d-metals from Hf to Au. DOS calculations at zero pressure were performed for the structures, experimentally proven to be stable at ambient pressure. The results are shown in Fig.\ref{fig:fcc}~(left). DOS under pressure was calculated for the structures observed at highest pressures reported for the respective elements experimentally. In fact, only two 5d-metals Hf and Au are unambiguously known to undergo a structural transformation under pressure.  Starting with Hf, we see that 4f and 5p levels are far from each other. However, with the increase of the atomic number (Z), 4f and 5p levels both shift towards higher binding energies. Importantly, 4f levels shift faster than 5p states. Thus 4f levels "outrun" 5p levels. In particular, $4f_{5/2}$ levels cross $5p_{3/2}$ already in Re and finally in Pt both $4f_{5/2}$ and $4f_{7/2}$ electrons lay below 5p electrons. 

\begin{table}[ht]
\caption {Data used for pressure estimations. Values with the superscripts are taken from the corresponding experimental studies. The rest is calculated in this study by fitting the equation of state with Vinet equation.}
\begin{tabular}{c|l*{9}{c}r}  \hline \hline   & \multicolumn{2}{c}{ Hf}  & Ta & W & Re & Os  & Ir  & Pt      & \multicolumn{2}{c}{Au}      \\ 
  
&bcc*&hcp&bcc&bcc&hcp&hcp&fcc&fcc&fcc&hcp*\\  
\hline
V$_0$        & 20.0 &  22.5   & 18.0     & 15.8      & 14.5     & 14.0    &14.2   &  15.1    & 17.1    & 18.1  \\ 
V$_P$        & 9.5  &  10.6   & 9.2      & 7.5       & 7.5      & 8.8     &12.1   &  7.2     & 7.1     & 9.2\\
B$_0$        & 114  & 113$^a$ & 195$^b$  & 312$^c$   & 353$^d$  & 399$^e$ &306$^f$&  273$^g$ & 167$^h$ & 172\\ 
B$_0^\prime$ & 3.7  & 3.3$^a$ & 3.4$^b$  & 4.3$^c$  & 4.6$^d$  & 4.0$^e$&6.8$^f$&  5.2$^g$ & 5.9$^h$ & 3.7\\
\hline  
\end{tabular}
\begin{flushleft}
*Bulk modulus and its derivative were calculated.
Volume per one atom is in $\AA^3$; bulk modulus is in GPa;\\
$^a$Reference \cite{Hrubiak2012}; 
$^b$Reference \cite{Cynn1999};
$^c$Reference \cite{He2006}; 
$^d$Reference \cite{Anzellini2014};
$^e$Reference \cite{Igor2015};
$^f$Reference \cite{Cerenius2000};
$^g$Reference  \cite{Matsui2009};
$^h$Reference \cite{Bercegeay2005};
\end{flushleft}
\end{table}

Fig.\ref{fig:fcc}~(right) demonstrates DOS of 5d metals upon compression. We provide DOS for structures which are known from experiment to be stable at high pressure. In fact, the crystal structure has very little effect on the core level crossing transitions, as will be discussed below.  Applied pressure causes broadening of p and s levels, so that they form rather broad bands due to overlap of the wave-functions. As it was shown in \cite{Igor2015} bulk modulus and its derivative calculated from Birch-Murnaghan fitting theoretically may give very inaccurate pressure estimate for the highly reduced volumes. Thus, for pressure estimation we used experimental bulk moduli and their pressure derivatives for cases where there were experimental measurements as shown (see Table I). 
 In Hf, Ta, W broadening of 5p levels due to pressure does not lead to overlapping of 4f and 5p levels even at highest pressures considered in this study. On the contrary, in Re $4f_{5/2}$ and $5p_{3/2}$ are overlapping at zero pressure and compression just causes a broadening of 5p levels. Since hcp Hf is unstable under high pressure and transforms first to $\omega$ phase, and then to bcc phase, we provide DOS for hcp and bcc structures for approximately the same volume reduction. One can clearly see that the structure does not have strong effect on the behavior of core-levels energies under pressure. We observed CLC of $4f_{7/2}$ and $5p_{3/2}$  in fcc Os at the same pressure as in \cite{Igor2015}. In Ir $4f_{5/2}$ and $5p_{1/2}$ states are very close, thus relatively small compression up to 80~GPa results in a CLC. These pressures are substantially lower then in Os, which makes it Ir extremely interesting in terms of a detailed experimental study of the CLC transition. In Pt 4f level lay below 5p and the distance between them is rather big. Therefore CLC  transition in Pt occurs at pressure above 1.5~TPa. Our calculations suggest that one should not expect to see a CLC in Au at any realistic pressures for either fcc or hcp structures. Analysis of the whole series of DOS for 5d-metals shows that in addition to broadening pressure causes shifts of all the core states towards higher binding energies and the shift is more pronounced for higher values of Z.  It is important to stress that the phenomenon of core-level crossing is strongly linked to the fact that f-level shift faster towards higher binding energies then p-level with increasing atomic number. Let us explain this effect.

\begin{figure}[h!]
\includegraphics[width=0.45\textwidth]{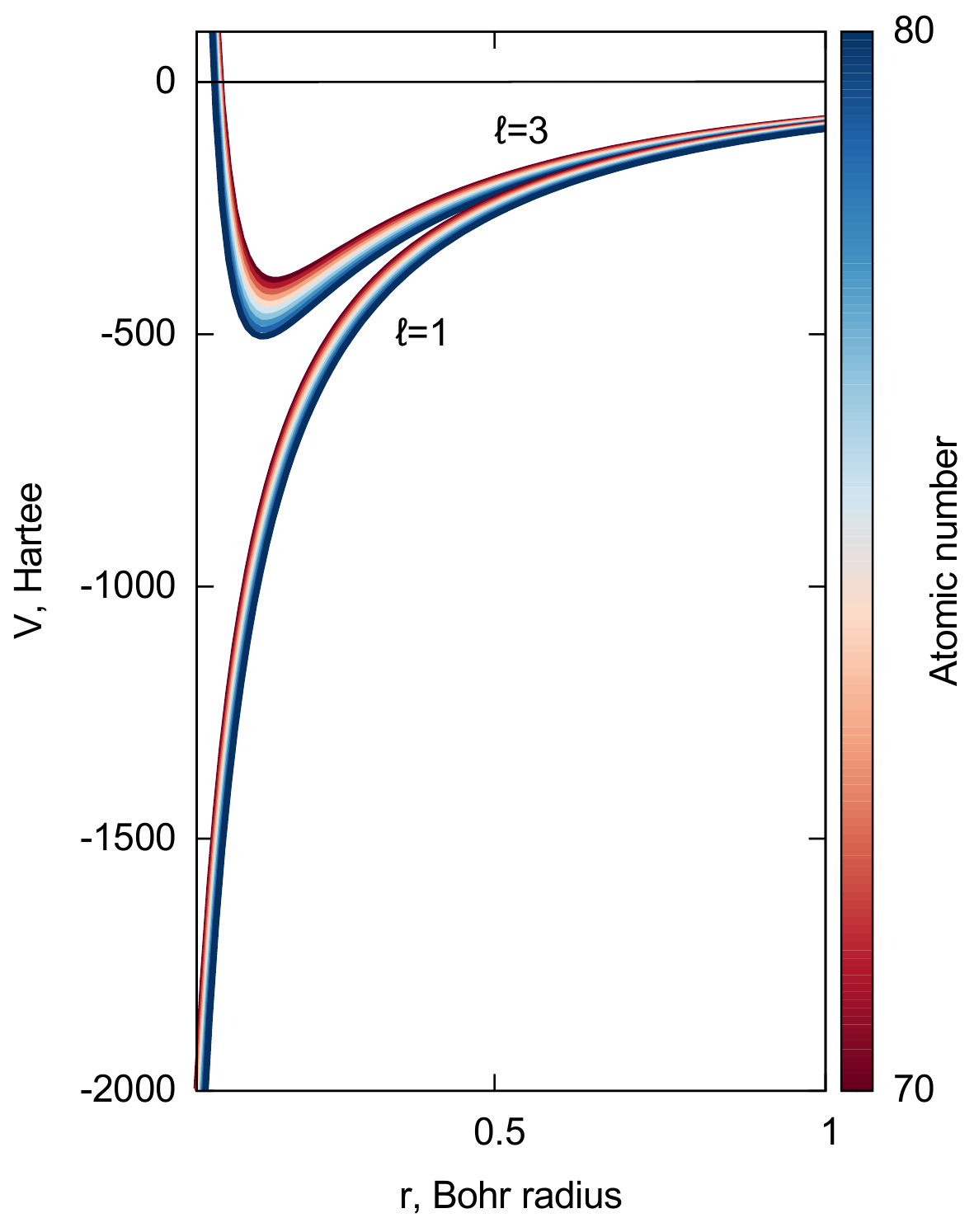}
\caption{Thomas-Fermi potentials for orbital numbers l=1 (p-electrons) and l=3 (f-electrons). Potentials are plotted for series of Z form 70 to 80. The color of the line corresponds to Z, as shown on the right palette.} 
\label{fig:potentials}
\end{figure}
\begin{figure*}[]
\begin{flushleft}

  \begin{tabular}{@{}cc@{}}
    \includegraphics[width=0.45\textwidth]{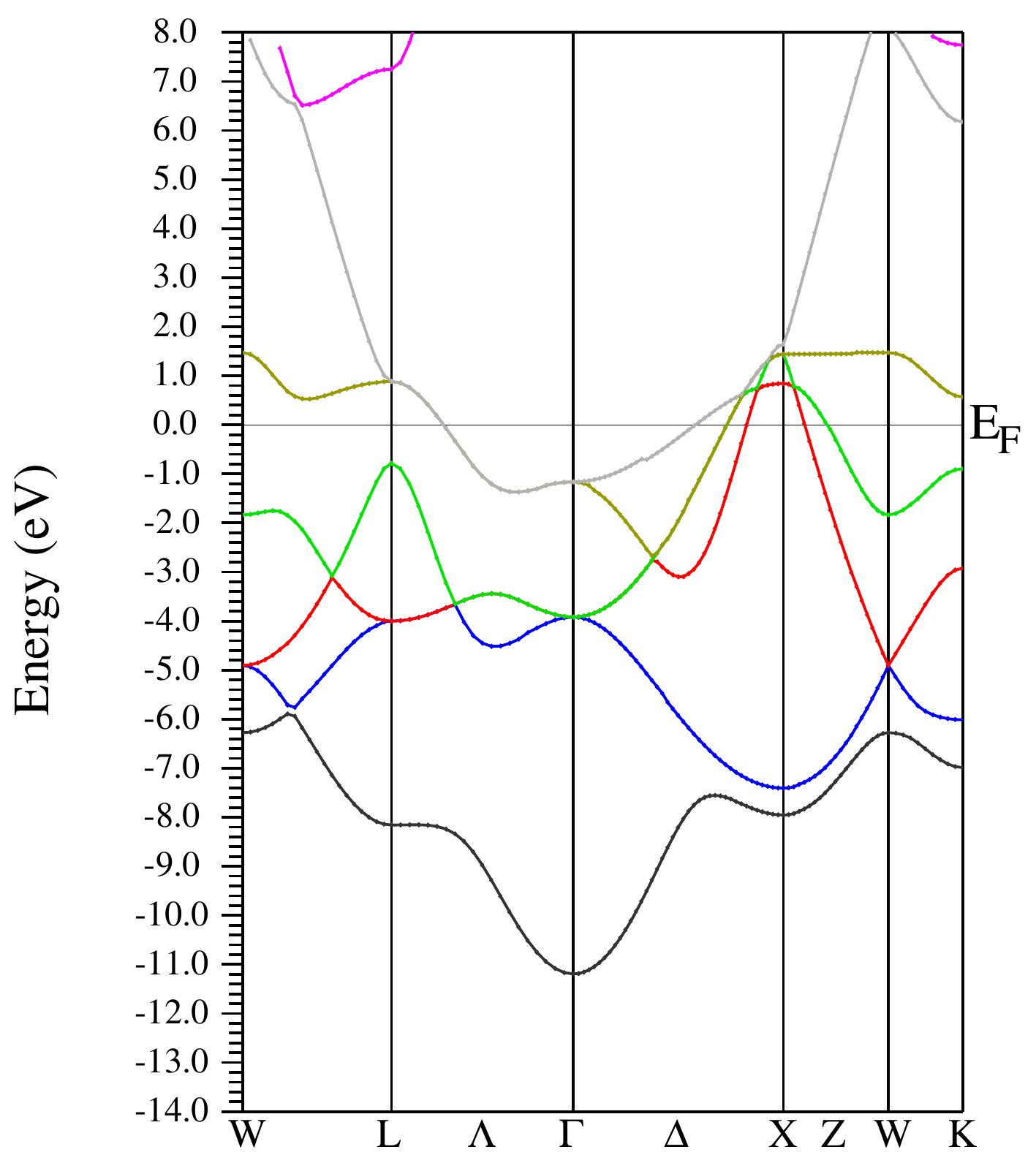} &
    \includegraphics[width=0.45\textwidth]{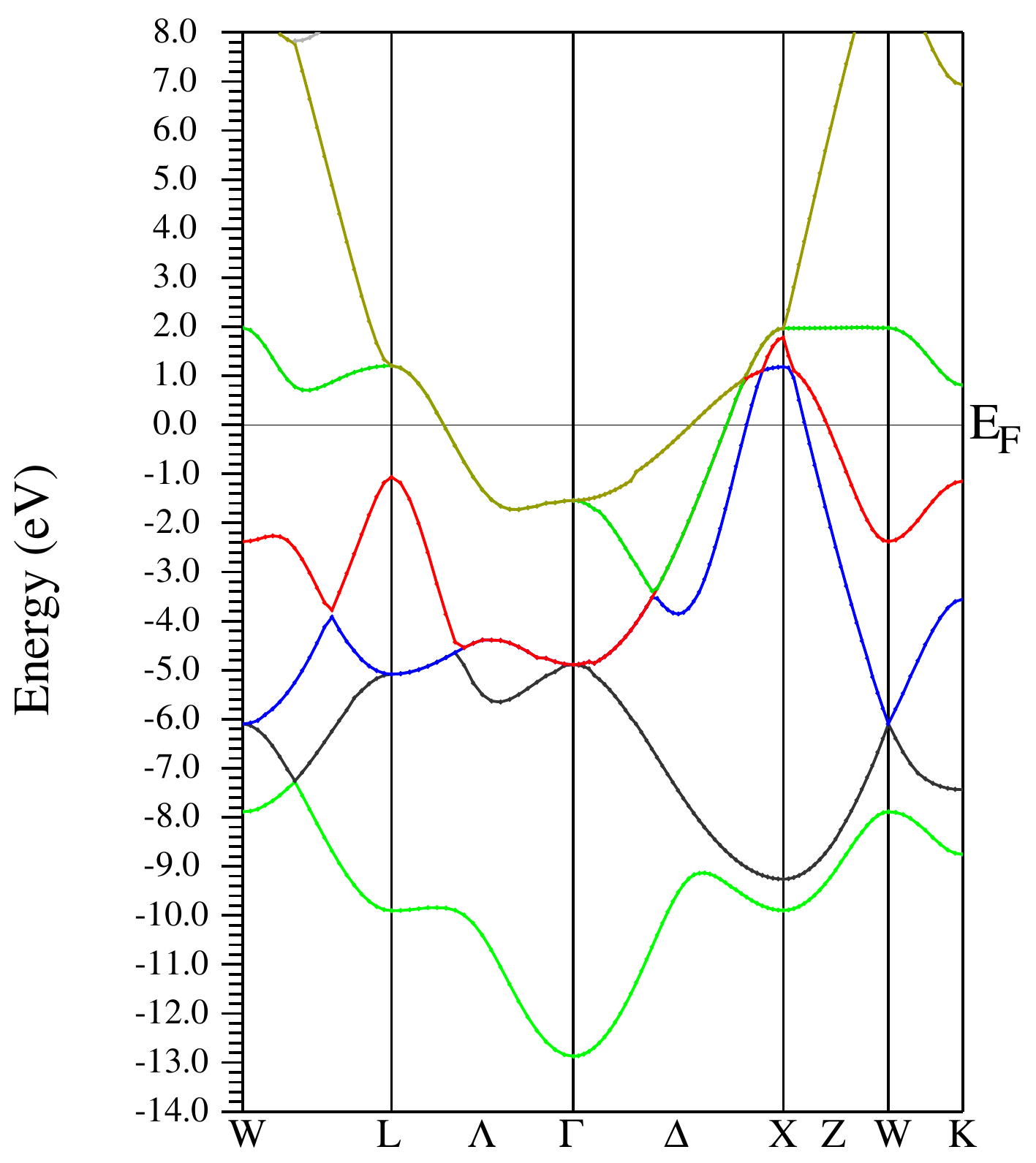}  \\

  \end{tabular}
  \caption{(left) Band structure of iridium at ambient pressure; (right) band structure of iridium under pressure of 80~GPa;}
\label{fig:band}
\end{flushleft}
\end{figure*}
In the early work M. Goeppert Mayer \cite{Mayer1941} demonstrated how atomic collapse appears in the rare-earth elements. She showed that the effective f-electrons potential suffers drastic changes with the increase of atomic number Z. 
In order to understand the behavior of f and p-levels in 5d-metals we analyzed the atomic potential behavior with the change of Z in the Thomas-Fermi approximation for the effective potential: 
\begin{equation}
V=-\frac{e^2}{r}[1+(Z-1)\phi(r / \mu)]+\frac{h^2}{8\pi^2 m}\frac{l(l+1)}{r^2}
\end{equation}  
 
The approximation suggested by Tietz \cite{Tietz1956} was used for the Thomas-Fermi function $\phi(r / \mu)$. 
\begin{equation}
\phi (x) = \frac{1}{(1+\alpha x)^2} 
\end{equation}  
where $x = r/a$, $a = 0.88534 Z^{-1/3}$ and $\alpha=0.53625$ \\
This approximation is believed to reproduce the shifts properly enough to distinguish the trends \cite{Mayer1941}.

Fig.\ref{fig:potentials} shows how potential V changes with atomic number. We considered series from Z=70 to 80. It is clear that the effective potential for f-electrons is more sensitive to Z. Indeed, and with the increase of Z both p and f-potentials become deeper but the f-potential shifts faster and f-levels will at some point outrun the p-levels.  This behavior is clearly seen in our first-principles results in Fig. 1. Two other effects seen in Fig. 1 are the pressure induced broadening, strongly pronounced for the p-levels, and a slight pressure induced up-shift of the states in energy, which is an effect of the higher spatial localization of the corresponding potential walls upon the decrease of the interatomic distances. These two effects lead to a crossing of the core levels for elements, in which they turned out to be sufficiently close to each other at ambient pressure.

Let us finally consider two elements, Ir and Re, where our calculations predict the CLC transition at relatively low pressure, and analyze the experimental information available in literature. In the fcc metals, like Ir, the CLC transition cannot be searched from the changes of c/a lattice parameters ratio, in contrast to hcp Os \cite{Igor2015}]. However, electronic transitions may lead to the changes of compressibility (such possibility has been discussed for Os \cite{Igor2015}). X-ray diffraction experiments performed by Cerenius and Dubrovinsky \cite{Cerenius2000} on fcc iridium showed distortion of the structure by the appearance of the additional diffraction peaks at the pressure exceeding 59~GPa, which was attempted to be explained by the formation of a superlattice. However, the superstructures were not observed in other experiments or in theoretical calculations. Moreover, a possible change of compressibility as a function of pressure was observed in recent experiments on Ir-rich Ir-Os alloy \cite{Yusenko2015}. All this points to a possible electronic transitions in these systems.  In Fig.\ref{fig:band} we show the band structure of Ir at P=0 GPa and at P=80 GPa, covering the pressure range of interest for this element. One can see that the presented bandstructure does not point to any Fermi-surface topology change and therefore the behavior can not be explained by an ETT.  On the other hand, in our calculations the CLC has been clearly shown at the pressure of 80~GPa, and it may give an explanation for the observed peculiarity. In fact, this situation is remarkably similar to the one, seen in Os at P$\approx$400 GPa \cite{Igor2015}. Of course, new experimental studies should be carried out for Ir.  On the other hand, a possibility of the structural transition in this  metal can complicate the task. For that reason, CLC could be verified experimentally in alloy systems, for example in Ir-Os alloy. Considering Re, the CLC transition is predicted to occur at already at ambient pressure. In this respect, it is interesting to point out that the results of investigations of the equation of state of Re are not totally consistent: while reported bulk moduli \cite{Anzellini2014,Igor2015,loz91} are basically the same within the uncertainty of the measurements, the reported pressure derivatives of the bulk moduli are substantially different. Therefore, Re is another interesting candidate for the studies of the CLC transition. Moreover, because we predict it to occur at ambient pressure, spectroscopic methods can be employed as well.

In summary, we have investigated the pressure dependence of the electronic structure of 5d metals from Hf to Au. We predict  that the newly discovered core-level crossing transition in Os also appears in Re, Ir, and Pt. We identify mechanisms that lead to the CLC transitions in the 5d-metals, which have many common features to the atomic collapse in the rare-earth elements.  The understanding of the CLC obtained in the present work allows us to expect to finding this phenomenon in variety of different systems. The fact that the CLC is predicted for metals used as pressure standards indicates that further investigations of this phenomenon are important.
 
The work was financially supported by the Knut and Alice
Wallenberg Foundation through Grant No. 2012.0083, the Swedish Government Strategic Research Area Grant Swedish e-Science Research Centre (SeRC) and “Strong Field Physics and New States of    Matter” from  Knut and Alice Wallenbergs Foundation.  I.A.A.
is grateful for the support provided by the Swedish Foundation
for Strategic Research (SSF) program SRL Grant No. 10-0026.
The support by the grant from the Ministry of Education and
Science of the Russian Federation (Grant No. 14.Y26.31.0005)
is gratefully acknowledged. The calculations were performed
on resources provided by the Swedish National Infrastructure
for Computing (SNIC) at the National Supercomputer Center
(NSC).

\bibliography{draft}{}

\end{document}